\documentclass[amsmath,amssymb,prx,hyperlink,twocolumn,superscriptaddress]{revtex4-1}

	\usepackage{graphicx}
	\usepackage{soul}
	\usepackage[colorlinks=true,citecolor=blue,linkcolor=magenta]{hyperref}
	\usepackage[usenames]{color}
	\usepackage{amsfonts}
	\usepackage{color}
	\usepackage{booktabs}
	\usepackage{multirow}
	\usepackage{float}
    \usepackage{dcolumn}
	
\usepackage{times}
\begin{document}

\title{Experimental Test of Irreducible Four-Qubit Greenberger-Horne-Zeilinger Paradox}

\author{Zu-En Su}
\affiliation{Hefei National Laboratory for Physical Sciences at Microscale and Department of Modern Physics, University of Science and Technology of China, Hefei, Anhui 230026, China}
\affiliation{CAS Centre for Excellence and Synergetic Innovation Centre in Quantum Information and Quantum Physics, University of Science and Technology of China, Hefei, Anhui 230026, China}

\author{Wei-Dong Tang}
\affiliation{Key Laboratory of Quantum Information and Quantum Optoelectronic Devices, Shanxi Province, and Department of Applied Physics of Xi'an Jiaotong University, Xi'an 710049, China}

\author{Dian Wu}
\affiliation{Hefei National Laboratory for Physical Sciences at Microscale and Department of Modern Physics, University of Science and Technology of China, Hefei, Anhui 230026, China}
\affiliation{CAS Centre for Excellence and Synergetic Innovation Centre in Quantum Information and Quantum Physics, University of Science and Technology of China, Hefei, Anhui 230026, China}

\author{Xin-Dong Cai}
\affiliation{Hefei National Laboratory for Physical Sciences at Microscale and Department of Modern Physics, University of Science and Technology of China, Hefei, Anhui 230026, China}
\affiliation{CAS Centre for Excellence and Synergetic Innovation Centre in Quantum Information and Quantum Physics, University of Science and Technology of China, Hefei, Anhui 230026, China}

\author{Tao Yang}
\affiliation{Hefei National Laboratory for Physical Sciences at Microscale and Department of Modern Physics, University of Science and Technology of China, Hefei, Anhui 230026, China}
\affiliation{CAS Centre for Excellence and Synergetic Innovation Centre in Quantum Information and Quantum Physics, University of Science and Technology of China, Hefei, Anhui 230026, China}

\author{Li Li}
\affiliation{Hefei National Laboratory for Physical Sciences at Microscale and Department of Modern Physics, University of Science and Technology of China, Hefei, Anhui 230026, China}
\affiliation{CAS Centre for Excellence and Synergetic Innovation Centre in Quantum Information and Quantum Physics, University of Science and Technology of China, Hefei, Anhui 230026, China}

\author{Nai-Le Liu}
\affiliation{Hefei National Laboratory for Physical Sciences at Microscale and Department of Modern Physics, University of Science and Technology of China, Hefei, Anhui 230026, China}
\affiliation{CAS Centre for Excellence and Synergetic Innovation Centre in Quantum Information and Quantum Physics, University of Science and Technology of China, Hefei, Anhui 230026, China}

\author{Chao-Yang Lu}
\affiliation{Hefei National Laboratory for Physical Sciences at Microscale and Department of Modern Physics, University of Science and Technology of China, Hefei, Anhui 230026, China}
\affiliation{CAS Centre for Excellence and Synergetic Innovation Centre in Quantum Information and Quantum Physics, University of Science and Technology of China, Hefei, Anhui 230026, China}

\author{Marek {\.Z}ukowski}
\affiliation{Institute for Theoretical Physics and Astrophysics, Faculty of Mathematics and   Informatics, University of Gdansk, Wita Stwosza 57, 80-308 Gdansk, Poland}

\author{Jian-Wei Pan}
\affiliation{Hefei National Laboratory for Physical Sciences at Microscale and Department of Modern Physics, University of Science and Technology of China, Hefei, Anhui 230026, China}
\affiliation{CAS Centre for Excellence and Synergetic Innovation Centre in Quantum Information and Quantum Physics, University of Science and Technology of China, Hefei, Anhui 230026, China}

\date{\today}

\begin{abstract}
Bell's theorem shows a profound contradiction between local realism and quantum mechanics on the level of statistical predictions. It does not involve directly Einstein-Podolsky-Rosen (EPR) correlations. The paradox of Greenberger-Horne-Zeilinger (GHZ) disproves directly the concept of EPR elements of reality, based on the EPR correlations, in an all-versus-nothing way. A three-qubit experimental demonstration of the GHZ paradox was achieved nearly twenty years ago, and followed by demonstrations for more qubits. Still, the GHZ contradictions underlying the tests can be reduced to three-qubit one. We show an irreducible four-qubit GHZ paradox, and report its experimental demonstration. The reducibility loophole is closed. The bound of a three-setting per party Bell-GHZ inequality is violated by $7\sigma$. The fidelity of the GHZ state was around $81\%$, and an entanglement witness reveals a violation of the separability threshold by $19\sigma$.
\end{abstract}

\maketitle

The EPR trio tried to argue that under quantum statistical predictions is hidden a local realistic description \cite{Einstein1935}. To this end they used the perfect correlations which are possible between pairs of entangled systems. Their attempt was disproved by Bell \cite{Bell1964}. Local realistic models cannot exits. After that many an experiment showed the thesis of Bell's theorem holds also for correlations observed in laboratories, see \cite{Pan2012} and \cite{brunner2014}. The final loophole-free experimental tests were done in 2015, \cite{Hensen2015} and \cite{Giustina2015,*Shalm2015}. However, Bell's theorem addresses statistical averages, and does not involve directly perfect EPR correlations, which (seemingly) suggest existence of (local) EPR elements of reality. In 1989, for more than two particles, Greenberger, Horne, and Zeilinger (GHZ) found  direct contradictions between quantum mechanics and the concept of elements of reality in \cite{Greenberger1989,*Greenberger1990}. The logic of their proof involved only perfect correlations, and was of an all-versus-nothing kind. In theory, not a single particle triple can be endowed with elements of reality. Such a striking paradox has sparked a widespread interest \cite{Bouwmeester2000}. The GHZ theorem has been extended to other systems and states, such as two-partite hyper-entangled states \cite{Cabello2001,*Chen2003,*Cinelli2005,*Yang2005,*Cabello2005,*Vallone2007}, code words \cite{DiVincenzo1997}, cluster states \cite{Scarani2005,*Walther2005,*Zhou2008,*Zhang2016}, graph states \cite{Tang2013a}, etc.

\begin{figure}[b]
\includegraphics[width=0.9\columnwidth]{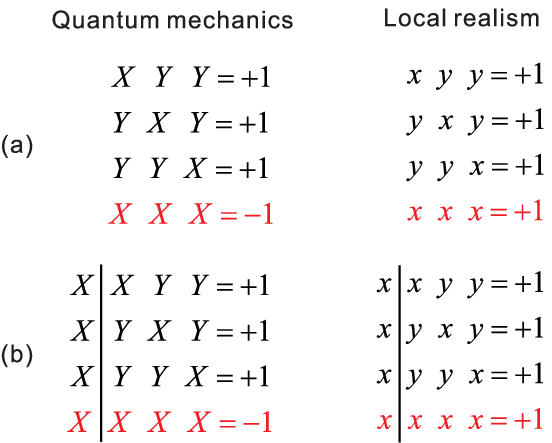}
\caption{\label{fig:ghz} Examples of irreducible and reducible GHZ paradoxes. (a) Irreducible GHZ paradox of three-qubit GHZ state \cite{Mermin1990a,Pan2000}. (b) Reducible GHZ paradox of four-qubit GHZ state \cite{Greenberger1989,Greenberger1990,Zhao2003}. $X$ and $Y$ are quantum predictions of Pauli measurement $\hat X$ and $\hat Y$, while $x$ and $y$ are corresponding predictions of local realism. The predictions in red (last line of each paradox) show an algebraic/logical contradiction.}
\end{figure}

Many efforts have been made to generalize the GHZ theorem to multipartite systems \cite{Cerf2002,Lee2006,Ryu2013,*Ryu2014,*Lawrence2014}. However, multipartite quantum entanglement displays a much richer and more complex structure than the tripartite case \cite{Mermin1990b}. Some constructed GHZ-type paradoxes may be explained by a hybrid local-nonlocal hidden variable model \cite{Cerf2002}, and the results can be reproduced by some biseparable entangled states. For instance, the original GHZ paradox for four qubits \cite{Greenberger1989,*Greenberger1990} actually can be reduced to the three qubits case \cite{Mermin1990b}. Another example was proof and experiment for four photons GHZ states \cite{Zhao2003}, see Fig. \ref{fig:ghz}. This holds also for four qubits cluster states \cite{Scarani2005,*Walther2005,*Zhou2008,*Zhang2016} and the most recent test of six photons GHZ states \cite{Zhang2015} (see Table I in the Supplemental Material \cite{SupplementalMaterial}).

Until recently irreducible multipartite GHZ paradoxes were known to for qubit systems which involve an odd number of particles \cite{Cerf2002,Lee2006,Ryu2013,*Ryu2014,Lawrence2014}. A set of observables which lead to a GHZ paradox must satisfy the following two conditions: (i) the observables, commuting or not, should share a common eigenstate (i.e., they must be ``concurrent"); (ii) no local observable appears only once in relations between elements of reality which jointly lead to a contradiction. In 2006, Lee \textit{et al.} \cite{Lee2006} put forward the first condition and presented the concept of concurrent observables. Recently, with the second condition, Tang \textit{et al.} \cite{Tang2013b} presented irreducible GHZ paradoxes for $d$ dimensional subsystems, which involve more than two measurement settings.

The additional motivation for all that effort is that irreducible $n$ particle GHZ paradoxes can be used to in quantum protocols such as $n$ partner secret sharing and controlled cryptography schemes \cite{Zukowski1998,*Hillery1999,*Kempe1999}. They allow perfect quantum reduction of communication complexity for some computation problems with distributed data \cite{Cleve1997}, etc.

In this Letter, we report the first experimental test of an irreducible four-qubit GHZ paradox \cite{Tang2013a,Tang2013b}. Thus we move into the class of even-number of particles correlations, for which a direct irreducible generalization of the original scheme \cite{Greenberger1989,*Greenberger1990} was thought to be impossible. Additionally, we confirm the validity of the test by showing a violation by the data of a corresponding Bell-type inequality, and we report high negative values of a witness observable of four-photon entanglement.

\begin{figure}[b]
\includegraphics[width=3.4 in]{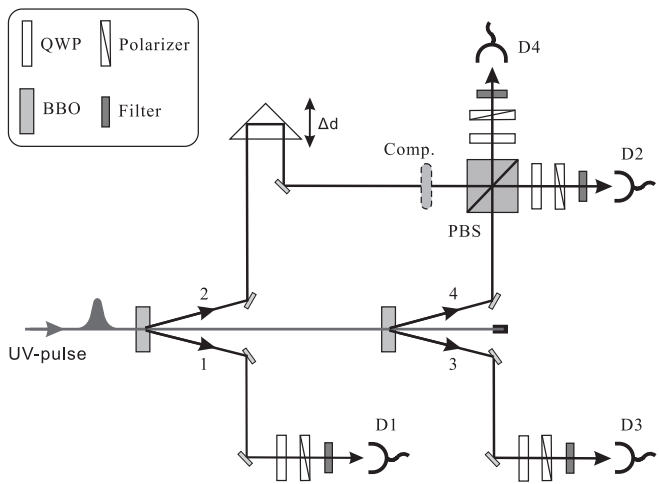}
\caption{\label{fig:setup} Experimental setup. Ultraviolet (UV) laser pulses pump two beta-barium borate (BBO) crystals. The instances in which top two pairs of entangled photons emerge, each from a different source, are used to produce the required four photon interference (this is signaled by detections in detectors D1 and D3). The UV laser which we use has a central wavelength of 394 nm and pulse duration of 120 fs, with a repetition rate of 76 MHz. The average pump power is 450 mW. Fine adjustments of the delays between path 2 and path 4 are made by translation stages \(\Delta d\) . We insert a compensator (Comp.) in front of the polarizing beam splitter (PBS) to counter the additional phase shifts of the PBS. To make the photons from the two sources indistinguishable at the detectors D2 and D4 behind the PBS, all photons are spectrally filtered (bandwidth 3.2 nm), in accordance with refs. \cite{Zukowski1993,*Zukowski1995}. The detection is made by fiber-coupled silicon single-photon detectors, whose total collection and detection efficiency is about 14\%. The four photon coincidence events are registered by a programmable multichannel coincidence unit. Quarter-wave plates (QWP) and polarizers are used to set the basis for the measurements of polarizations of the photons.}
\end{figure}

{\em The GHZ state and its laboratory generation.---}
We prepare four-photon GHZ state using two source spontaneous parametric down-conversion of type-II \cite{Kwiat1995}, which gives us two-entangled-pairs emissions, and the state construction methods of multi-photon interferometry \cite{Pan2012}. As shown in Fig. \ref{fig:setup}, an ultraviolet pulse laser passes through two beta-barium borate crystals. This gives a non-vanishing probability to produce two photon pairs, each with an entangled state of polarizations $(\left| H \right\rangle \left| H \right\rangle  + \left| V \right\rangle \left| V \right\rangle )/\sqrt 2$, where $\left| H \right\rangle$ ($\left| V \right\rangle$) denotes horizontal (vertical) linear polarization state \cite{Kwiat1995}. One photon from each pair is then directed to an input of a polarizing beam splitter (PBS), at which they arrive simultaneously. If total indistinguishability of sources of the photons which emerge the PBS is enforced, since the PBS transmits horizontal and reflects vertical polarization, coincidence detection between the two outputs of the PBS implies that either both photons 2 and 4 are horizontally polarized or both vertically polarized, and thus projects the four photons state into a coherent superposition giving the GHZ state:
\begin{equation}
\frac{1}{{\sqrt 2 }}({\left| H \right\rangle _1}{\left| H \right\rangle _2}{\left| H \right\rangle _3}{\left| H \right\rangle _4} + {\left| V \right\rangle _1}{\left| V \right\rangle _2}{\left| V \right\rangle _3}{\left| V \right\rangle _4}).
\label{eq:state}
\end{equation}
In our setup the four photons coincidence rate which was consistent with components in (\ref{eq:state}) is about 6.8 Hz. The ratio of such `good' events to the undesired ones, such as related with the other possible components like ${\left| H \right\rangle _1}{\left| H \right\rangle _2}{\left| H \right\rangle _3}{\left| V \right\rangle _4}$, etc., was better than $8:1$. The four particle interferometric contrast (visibility)  was 0.733\( \pm \)0.039, in the basis $(\left| H \right\rangle  \pm \left| V \right\rangle)/\sqrt 2$. Hence, the coherent superposition in (\ref{eq:state}) was observed (with a small mixture of noise).

{\em Irrreducible GHZ paradox.---}
The quantum predictions for the state (\ref{eq:state}) allow one to formulate an irreducible four-qubit GHZ paradox. Reducible GHZ paradoxes \cite{Zhao2003,Zhang2015} use only two settings, see Fig.~\ref{fig:ghz}(b). Here we use three settings for each observation station, to get a four particle irreducible GHZ contradiction \cite{Tang2013b}. The three local observables are
\begin{equation}
\hat{X}(\theta) = {e^{i\theta}}\left| H \right\rangle \left\langle V \right| + {e^{- i\theta}}\left| V \right\rangle \left\langle H \right|,
\label{eq:observable}
\end{equation}
where $\theta$ equals either $0$, or  $-\pi/4$, or $\pi/8$. Their eigenstates with eigenvalues $\pm 1$ are $(\left| H \right\rangle  \pm \left| V \right\rangle)/\sqrt 2 $, $ (\pm {e^{i\pi /4}}\left| H \right\rangle  + \left| V \right\rangle )/ \sqrt 2 $, and $ (\pm {e^{ - i\pi /8}}\left| H \right\rangle  + \left| V \right\rangle )/ \sqrt 2$, respectively. In the experiment, the measurement of observable $\hat{X}(0)$ corresponds to the analysis of \( \pm 45^\circ \) linear polarization. To measure observables $\hat{X}(- \pi/4)$ and $\hat{X}(\pi/8)$, a quarter wave plate (QWP) at \(45^\circ \) is inserted. This transforms the eigenstates into $67.5^\circ$/$ - 22.5^\circ$ and $33.75^\circ$/$ - 56.25^\circ$ linear polarization states, respectively. Thus, all three observables can be measured with a QWP and a polarizer, as shown in Fig. \ref{fig:setup}.

\begin{figure*}
\includegraphics[width=7.0 in]{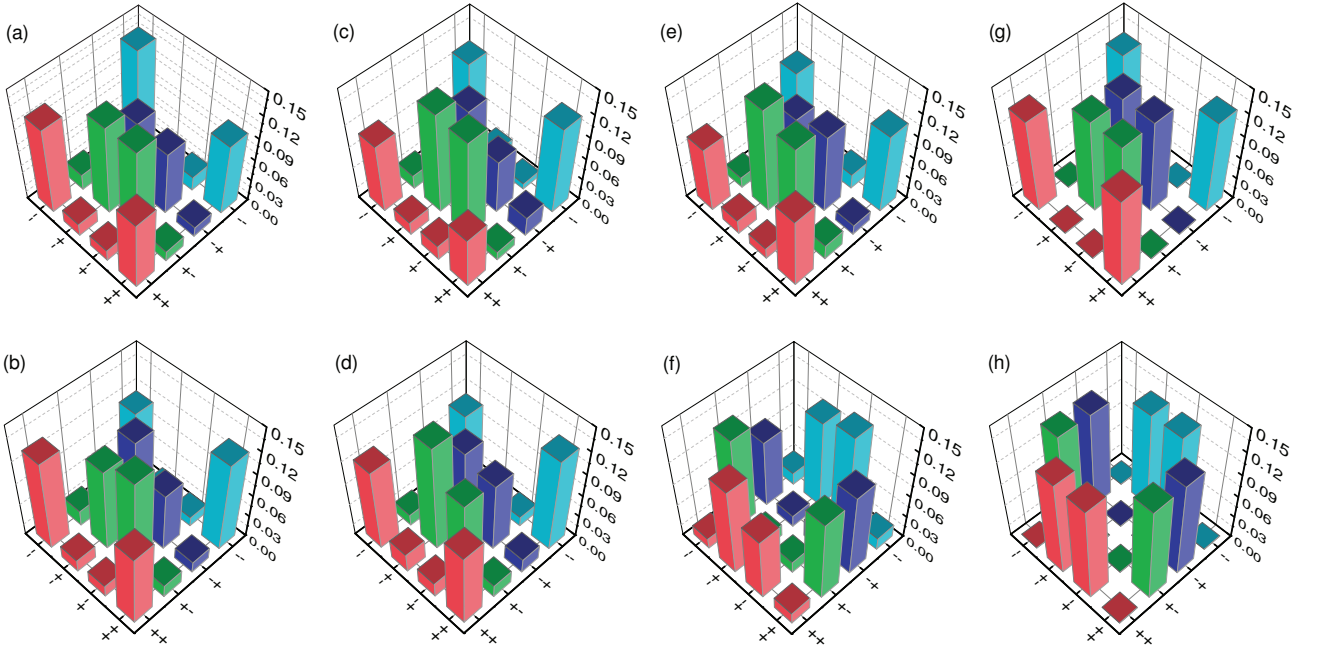}
\caption{\label{fig:results} (color online). Experimental estimates of probability distributions observed in the measurement arrangements related with the global observables  $\hat{X}^{I}$ (a), $\hat{X}^{II}$ (b), $\hat{X}^{III}$ (c), $\hat{X}^{IV}$ (d), $\hat{X}^{V}$ (e), $\hat{X}^{VI}$ (f). In (h) we show the ideal quantum prediction for measurement arrangement (f), while (g) represents the $100\%$ flipped situation which must arise in the case of local realism, if the first five situations (a-e) follow ideal quantum prediction (recall the 1=-1 contradiction). In the figure, $\pm$ stands for corresponding one-qubit eigenstates with eigenvalues $\pm 1$. The interferometric visibilities in situations (a)-(e) are 0.733\( \pm \)0.039, 0.730\( \pm \)0.043, 0.704\( \pm \)0.042, 0.711\( \pm \)0.041, and 0.727\( \pm \)0.043, respectively. The experimental results in (f) are in agreement with the quantum mechanics predictions (h) while in a strong conflict with local realism (g), and this is with a visibility of 0.733\( \pm \)0.042 or the interference in the situation (f). Visibility is defined as $({N_{\max }} - {N_{\min }})/({N_{\max }} + {N_{\min }})$, where ${N_{\max }}$ is the sum of eight ``tall" probabilities, and ${N_{\min }}$ is the sum of other eight ``short" probabilities. The integration time of each fourfold coincidence is 3 minutes. }
\end{figure*}

To show the GHZ paradox, first consider the following five measurements, i.e.
\begin{eqnarray}\label{eq:five}
&\hat{X}^{I} \equiv {\hat{X}_1}(0){\hat{X}_2}(0){\hat{X}_3}(0){\hat{X}_4}(0),&\nonumber\\
&\hat{X}^{II} \equiv {\hat{X}_1}(0){\hat{X}_2}( -\frac{\pi }{4}){\hat{X}_3}( \frac{\pi }{8}){\hat{X}_4}( \frac{\pi }{8}),&\nonumber\\
&\hat{X}^{III} \equiv {\hat{X}_1}( \frac{\pi }{8}){\hat{X}_2}(0){\hat{X}_3}(- \frac{\pi }{4}){\hat{X}_4}( \frac{\pi }{8}),&\\
&\hat{X}^{IV} \equiv {\hat{X}_1}( \frac{\pi }{8}){\hat{X}_2}( \frac{\pi }{8}){\hat{X}_3}(0){\hat{X}_4}( -\frac{\pi }{4}),&\nonumber\\
&\hat{X}^{V} \equiv {\hat{X}_1}( -\frac{\pi }{4}){\hat{X}_2}( \frac{\pi }{8}){\hat{X}_3}( \frac{\pi }{8}){\hat{X}_4}(0).&\nonumber
\end{eqnarray}
The state (\ref{eq:state}) is a common eigenstate of the five operators (\ref{eq:five}), with eigenvalue $+1$.

Assume that each photon carries EPR elements of reality for each of the three local observables $\hat{X}(0)$, $\hat{X}(-\pi/4)$, and $\hat{X}(\pi/8)$. For photon $i$, we denote the elements of reality by ${x_i(0)}$. The quantum prediction that for each of the five global observables $\hat{X}^{L}$, where $L=I,II, III, IV, V$ we always get $+1$, implies that for the elements of reality for the constituent local observables must satisfy
\begin{eqnarray}\label{EPR-5}
{x_1}(0){x_2}(0){x_3}(0){x_4}(0) = + 1,\nonumber\\
{x_1}(0){x_2}( -\frac{\pi }{4}){x_3}( \frac{\pi }{8}){x_4}( \frac{\pi }{8}) = + 1,\nonumber\\
{x_1}( \frac{\pi }{8}){x_2}(0){x_3}( -\frac{\pi }{4}){x_4}( \frac{\pi }{8}) = + 1,\\
{x_1}( \frac{\pi }{8}){x_2}( \frac{\pi }{8}){x_3}(0){x_4}( -\frac{\pi }{4}) = + 1,\nonumber\\
{x_1}( -\frac{\pi }{4}){x_2}( \frac{\pi }{8}){x_3}( \frac{\pi }{8}){x_4}(0) = + 1.\nonumber
\end{eqnarray}

Now let us consider the sixth global observable,
\begin{equation}
\hat{X}^{VI} \equiv {\hat{X}_1}( - \frac{\pi }{4}){\hat{X}_2}( - \frac{\pi }{4}){\hat{X}_3}(- \frac{\pi }{4}){\hat{X}_4}( -\frac{\pi }{4}).
\label{eq:six}
\end{equation}
The observable (\ref{eq:six}) has an eigenvalue $-1$ when acting on its eigenstate (\ref{eq:state}). Thus the local ERR elements of reality associated with $\hat{X}^{VI}$ must satisfy
\begin{equation}\label{EPR-6}
{x_1}( - \frac{\pi }{4}){x_2}( - \frac{\pi }{4}){x_3}( - \frac{\pi }{4}){x_4}( - \frac{\pi }{4}) =  - 1.
\end{equation}
However if one multiplies side by side all six relations of the EPR elements of reality, (\ref{EPR-5}) and (\ref{EPR-6}), since whatever is the value of the local phase $\phi$ one has $x_i(\phi)^2=1$, one gets an algebraic/logical contradiction $1=-1$. This disproves existence of EPR elements of reality, and the entire EPR method, as well as contradicts principles of local realism (equivalently, local causality).

The experimental results of measurement all the above six global observables [Fig. \ref{fig:results}] show that, the coincidences predicted by quantum mechanics occur, albeit with some noise. Local realistic description is excluded. The spurious events, of a small probability, are unavoidable experimental errors caused by high-order photon emissions and imperfections of the photonic components. If local realism could reproduce experimental results [Fig. \ref{fig:results}(f)], we have to assume that the spurious events in the earlier five experiments [Fig. \ref{fig:results}(a)-(e)] actually indicate a deviation from quantum mechanics \cite{Pan2000}. In such a case, the fraction of correct events in the experiment sixth global measurement arrangement [Fig. \ref{fig:results}(f)] can at most be equal to the sum of the fractions of all spurious events in the other five experiments [Fig. \ref{fig:results}(a)-(e)], that is, \(0.70 \pm 0.02\). However, we observed such terms with a fraction of \(0.87 \pm 0.02\) [Fig. \ref{fig:results}(f)], which violates threshold for a consistent local realistic model for all six situations by more than seven standard deviations. Thus, the correlations in the measurements invalidate any local realistic model for our GHZ entanglement, and are concurrent with the quantum predictions.

This conflict can also be seen in a violation of a suitable multi-setting Bell-type inequality \cite{Tang2013b} derivable directly from the GHZ paradox conditions. Obviously it differs from the optimal, two-setting-per-observer Bell's inequality for four-particle GHZ state, i.e. Mermin-Ardehali-Belinskii-Klyshko inequality \cite{Ardehali1992}, the violation of which was demonstrated by Zhao \textit{et al.} \cite{Zhao2003}. The three setting per party GHZ-Bell inequality reads
\begin{equation}
\left| B \right| \le 4,
\label{eq:inequality}
\end{equation}
where $B$ is a Bell-GHZ observable $\hat{B} = \hat{X}^{I} + \hat{X}^{II} + \hat{X}^{III} + \hat{X}^{IV} +\hat{X}^{V} - \hat{X}^{VI}$. Quantum mechanics predicts the maximal violation of the inequality (\ref{eq:inequality}) by a factor of 3/2, as one can saturate its algebraic bound of 6. Hence, the threshold visibility to violate the constraint (\ref{eq:inequality}) is given by 2/3 $\simeq$ 66.7\%. Note that this increases the experimental difficulty relative to the optimal two-setting per party Bell inequality for four-particle GHZ state, whose threshold visibility is 35.4\% \cite{Ardehali1992} (34.2\% for a three-setting one, however of a different kind than ours \cite{Zukowski1997}). The average visibility observed in the experiment for the state (\ref{eq:state}) is 72.3\% and thus exceeds the threshold. Substituting the experimental results [Fig.~\ref{fig:results}] into the left-hand side of inequality (\ref{eq:inequality}) gives
\begin{equation}
\left| B \right| = 4.34 \pm 0.04.
\label{eq:violation}
\end{equation}
The violation of the inequality (\ref{eq:inequality}) is by over 8.5 standard deviations. In this way, since the inequality involves only perfect (in the quantum case) EPR-GHZ correlations, with three different setting for each observer, its violation is an unambiguous evidence for the validity of our experimental test of an irreducible four-qubit GHZ paradox.

We also measured a suitably chosen entanglement witness \cite{Bourennane2004,*Guhne2007}, which is a measure indicating whether genuine multipartite entanglement was produced in the experiment. The measurements (see the Supplemental Material \cite{SupplementalMaterial}) yield the value of the entanglement witness of \( - 0.306 \pm 0.016\), which is negative by 19 standard deviations and thus proves the presence of genuine four-partite entanglement. From the expectation value of the witness, we can directly determine the obtained fidelity of the GHZ state as \(0.806 \pm 0.016\), which significantly exceeds the threshold of 50\% \cite{Guhne2009}.

In conclusion, we have demonstrated experimentally the first irreducible four-photon non-statistical conflict between quantum mechanics and local realism. Our results close the reducibility loophole and thus open a new way to testing genuine GHZ paradoxes with more photons, which will be helpful for designing new quantum protocols, like e.g. \cite{Zukowski1998,*Hillery1999,*Kempe1999}. Besides, the multi-setting approach can be used in studies of multipartite contextuality \cite{Tang2013a, Raeisi2015}. It should be pointed out that, just as all other GHZ experiments done so far, our experiment suffers from the loopholes of locality, freedom-of-choice and detection efficiency. However, recent breakthrough Bell tests have closed all these loopholes simultaneously in the two qubit case \cite{Giustina2015,*Shalm2015,Hensen2015}. Possible future experiments can combine these techniques and the methods used here to further study observability of GHZ correlations in the multipartite or multilevel systems \cite{Tang2013a,Tang2013b,Cerf2002,Lee2006,Ryu2013,*Ryu2014,Lawrence2014} without loopholes \cite{Erven2014}.

{\em Acknowledgments---} This work was supported by the National Natural Science Foundation of China, the Chinese Academy of Sciences, and the National Fundamental Research Program. WDT acknowledge financial support by the NNSF of China (Grant No. 11405120). M{\.Z} is supported by EU advanced grant QOLAPS, and COPERNICUS grant-award of DFG/FNP.

\section*{APPENDIX Supporting Materials}
\renewcommand{\thefigure}{S\arabic{figure}}
 \setcounter{figure}{0}
\renewcommand{\theequation}{S.\arabic{equation}}
 \setcounter{equation}{0}
 \renewcommand{\thesection}{S.\Roman{section}}
\setcounter{section}{0}

\begin{figure}[b]
\includegraphics[width=3.3 in]{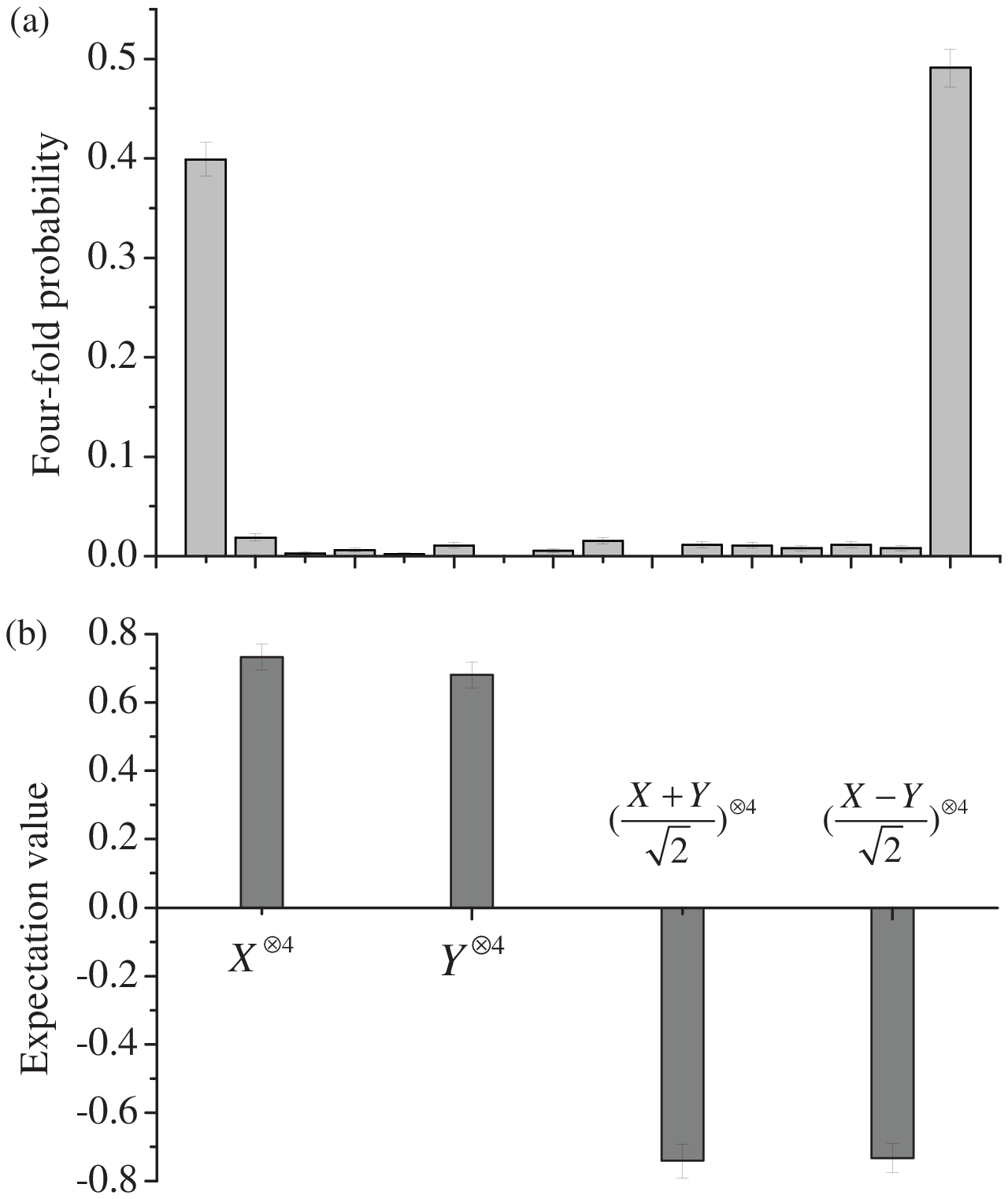}
\caption{\label{fig:fidelity} (color online). Measurement of witness and fidelity of the four-photon GHZ state. (a) The results of Pauli measurement \({Z^{ \otimes 4}}\) are listed. From left to right: $HHHH$, $HHHV$, $HHVH$, $HHVV$, $HVHH$, $HVHV$, $HVVH$, $HVVV$, $VHHH$, $VHHV$, $VHVH$, $VHVV$, $VVHH$, $VVHV$, $VVVH$, $VVVV$. Here $HHHH$ is short for the measuring eigenstate ${\left| H \right\rangle _1}{\left| H \right\rangle _2}{\left| H \right\rangle _3}{\left| H \right\rangle _4}$, etc. (b) Expectation values of other 4 observables are listed. Each of them is derived from a complete set of 16 four-fold coincidence events similar to (a). The error bars denote standard deviation, deduced from propagated Poissonian counting statistics of the raw detection events.}
\end{figure}

The entanglement witness of the four-qubit GHZ state used in the main text can be decomposed into a series of local projective measurements \cite{Guhne2007}, i.e.
\begin{equation}
{W_{GHZ}} = \frac{I}{2} - \left| GHZ  \right\rangle \left\langle GHZ  \right|
\end{equation},
where $I$ denotes the identity operator, and
\begin{equation}
\begin{split}
\left| GHZ  \right\rangle \left\langle GHZ  \right| = \frac{1}{2}((\left| H \right\rangle \left\langle H \right|)^{ \otimes 4} + (\left| V \right\rangle \left\langle V \right|)^{ \otimes 4}) \\
+ \frac{1}{8}({X^{ \otimes 4}} + {Y^{ \otimes 4}} - {(\frac{{X + Y}}{{\sqrt 2 }})^{ \otimes 4}} - {(\frac{{X - Y}}{{\sqrt 2 }})^{ \otimes 4}})
\end{split}
\end{equation},
where $X$, $Y$ are Pauli operators, and $\left| H \right\rangle$ ($\left| V \right\rangle$) denotes horizontal (vertical) linear polarizing state. Fig. \ref{fig:fidelity} shows the measurement results.

The fidelity between $\left| GHZ  \right\rangle$ and the prepared state ${\rho _{\exp }}$ is given by $F({\rho _{\exp }}) = \left\langle GHZ  \right|{\rho _{\exp }}\left| GHZ  \right\rangle$ \cite{Nielsen2010}.

\begin{table*}
\caption{\label{tab:table}
A reducible GHZ paradox can be reproduced with quantum states had a separable partition rather than the genuine multi-partite entangled state. Some examples of reducible GHZ paradoxes in qubit systems are listed below: four-qubit case of cluster state in a one-dimensional lattice \cite{Scarani2005,*Walther2005,*Zhou2008,*Zhang2016}, four-qubit case of GHZ state \cite{Zhao2003}, five-qubit case \cite{Pagonis1991}, six-qubit cases \cite{Pagonis1991,Zhang2015}, and eight-qubit cases \cite{Zhang2015}.}
\begin{ruledtabular}
\begin{tabular}{ccc}
Number of qubits&GHZ paradoxes&Bi-separable quantum states\\
\hline
4 & \(\begin{array}{c}
{Z_1}{X_2}{Y_3}{Y_4} =  + 1\\
{Z_1}{Y_2}{\rm{ }}{X_3}{Y_4} =  + 1\\
{I_1}{X_2}{X_3}{Z_4} =  + 1\\
{I_1}{Y_2}{Y_3}{Z_4} =  - 1
\end{array}\) & \(\begin{array}{c}
{\left| 0 \right\rangle _1} \otimes ({\left| 0 \right\rangle _2}{\left| 0 \right\rangle _3}{\left| 0 \right\rangle _4} + {\left| 0 \right\rangle _2}{\left| 0 \right\rangle _3}{\left| 1 \right\rangle _4} + {\left| 1 \right\rangle _2}{\left| 1 \right\rangle _3}{\left| 0 \right\rangle _4} - {\left| 1 \right\rangle _2}{\left| 1 \right\rangle _3}{\left| 1 \right\rangle _4})
\end{array}\)\\
\\
4 & \(\begin{array}{c}
{Z_1}{X_2}{X_3}{X_4} =  + 1\\
{Z_1}{Y_2}{\rm{ }}{X_3}{Y_4} =  + 1\\
{Z_1}{X_2}{Y_3}{Y_4} =  + 1\\
{Z_1}{Y_2}{Y_3}{X_4} =  - 1
\end{array}\) & \(\begin{array}{c}
{\left| 0 \right\rangle _1}
 \otimes ({\left| 0 \right\rangle _2}{\left| 0 \right\rangle _3}{\left| 1 \right\rangle _4} + i{\left| 1 \right\rangle _2}{\left| 1 \right\rangle _3}{\left| 0 \right\rangle _4})
\end{array}\)\\
5 & \(\begin{array}{c}
\\
{X_1}{X_2}{X_3}{X_4}{X_5} =  + 1\\
{X_1}{Y_2}{Y_3}{Y_4}{Y_5} =  + 1\\
{Y_1}{X_2}{Y_3}{Y_4}{Y_5} =  + 1\\
{Y_1}{Y_2}{X_3}{X_4}{X_5} =  - 1
\end{array}\) & \(\begin{array}{c}
({\left| 0 \right\rangle _1}{\left| 0 \right\rangle _2}{\left| 0 \right\rangle _3} + {\left| 1 \right\rangle _1}{\left| 1 \right\rangle _2}{\left| 1 \right\rangle _3})\\
 \otimes ({\left| 0 \right\rangle _4}{\left| 0 \right\rangle _5} + {\left| 1 \right\rangle _4}{\left| 1 \right\rangle _5})
\end{array}\)\\
6 & \(\begin{array}{c}
\\
{X_1}{Y_2}{Y_3}{Y_4}{Y_5}{Y_6} =  + 1\\
{X_1}{Y_2}{Y_3}{Y_4}{Y_5}{Y_6} =  + 1\\
{Y_1}{X_2}{Y_3}{Y_4}{Y_5}{Y_6} =  + 1\\
{Y_1}{Y_2}{X_3}{Y_4}{Y_5}{Y_6} =  + 1\\
{Y_1}{X_2}{X_3}{X_4}{X_5}{X_6} =  + 1\\
{Y_1}{Y_2}{Y_3}{X_4}{X_5}{X_6} =  - 1
\end{array}\) & \(\begin{array}{c}
({\left| 0 \right\rangle _1}{\left| 0 \right\rangle _2}{\left| 0 \right\rangle _3}{\left| 0 \right\rangle _4} + i{\left| 1 \right\rangle _1}{\left| 1 \right\rangle _2}{\left| 1 \right\rangle _3}{\left| 1 \right\rangle _4})\\
 \otimes ({\left| 0 \right\rangle _5}{\left| 0 \right\rangle _6} + {\left| 1 \right\rangle _5}{\left| 1 \right\rangle _6})
\end{array}\)\\
6 & \(\begin{array}{c}
\\
{Y_1}{Y_2}{X_3}{X_4}{X_5}{X_6} =  - 1\\
{X_1}{Y_2}{Y_3}{X_4}{X_5}{X_6} =  - 1\\
{X_1}{Y_2}{X_3}{Y_4}{X_5}{X_6} =  - 1\\
{X_1}{X_2}{X_3}{Y_4}{Y_5}{X_6} =  - 1\\
{X_1}{X_2}{Y_3}{X_4}{X_5}{Y_6} =  - 1\\
{Y_1}{Y_2}{X_3}{X_4}{Y_5}{Y_6} =  + 1
\end{array}\) & \(\begin{array}{c}
({\left| 0 \right\rangle _1}{\left| 0 \right\rangle _3}{\left| 0 \right\rangle _4} + i{\left| 1 \right\rangle _1}{\left| 1 \right\rangle _3}{\left| 1 \right\rangle _4})\\
 \otimes ({\left| 0 \right\rangle _2}{\left| 0 \right\rangle _5}{\left| 0 \right\rangle _6} + i{\left| 1 \right\rangle _2}{\left| 1 \right\rangle _5}{\left| 1 \right\rangle _6})
\end{array}\)\\
6 & \(\begin{array}{c}
\\
{Y_1}{Y_2}{X_3}{X_4}{X_5}{X_6} =  - 1\\
{Y_1}{X_2}{X_3}{X_4}{Y_5}{X_6} =  - 1\\
{X_1}{Y_2}{X_3}{Y_4}{X_5}{X_6} =  - 1\\
{X_1}{X_2}{Y_3}{X_4}{Y_5}{X_6} =  - 1\\
{Y_1}{Y_2}{Y_3}{Y_4}{Y_5}{Y_6} =  - 1\\
{Y_1}{Y_2}{X_3}{X_4}{Y_5}{Y_6} =  - 1
\end{array}\) & \(\begin{array}{c}
({\left| 0 \right\rangle _1}{\left| 0 \right\rangle _3}{\left| 0 \right\rangle _4} + i{\left| 1 \right\rangle _1}{\left| 1 \right\rangle _3}{\left| 1 \right\rangle _4})\\
 \otimes ({\left| 0 \right\rangle _2}{\left| 0 \right\rangle _5}{\left| 0 \right\rangle _6} + i{\left| 1 \right\rangle _2}{\left| 1 \right\rangle _5}{\left| 1 \right\rangle _6})
\end{array}\)\\
6 & \(\begin{array}{c}
\\
{X_1}{X_2}{Y_3}{Y_4}{X_5}{X_6} =  - 1\\
{Y_1}{X_2}{Y_3}{X_4}{X_5}{X_6} =  - 1\\
{X_1}{Y_2}{Y_3}{X_4}{X_5}{X_6} =  - 1\\
{X_1}{X_2}{Y_3}{X_4}{Y_5}{X_6} =  - 1\\
{X_1}{X_2}{X_3}{Y_4}{X_5}{Y_6} =  - 1\\
{Y_1}{Y_2}{X_3}{X_4}{Y_5}{Y_6} =  + 1
\end{array}\) & \(\begin{array}{c}
({\left| 0 \right\rangle _1}{\left| 0 \right\rangle _2}{\left| 0 \right\rangle _4}{\left| 0 \right\rangle _5} + i{\left| 1 \right\rangle _1}{\left| 1 \right\rangle _2}{\left| 1 \right\rangle _4}{\left| 1 \right\rangle _5})\\
 \otimes ({\left| 0 \right\rangle _3}{\left| 0 \right\rangle _6} + i{\left| 1 \right\rangle _3}{\left| 1 \right\rangle _6})
\end{array}\)\\
6 & \(\begin{array}{c}
\\
{Y_1}{X_2}{Y_3}{X_4}{X_5}{X_6} =  - 1\\
{Y_1}{Y_2}{X_3}{X_4}{X_5}{X_6} =  - 1\\
{Y_1}{X_2}{X_3}{X_4}{Y_5}{X_6} =  - 1\\
{X_1}{X_2}{Y_3}{Y_4}{X_5}{X_6} =  - 1\\
{X_1}{X_2}{X_3}{Y_4}{X_5}{Y_6} =  - 1\\
{Y_1}{Y_2}{X_3}{X_4}{Y_5}{Y_6} =  + 1
\end{array}\) & \(\begin{array}{c}
({\left| 0 \right\rangle _1}{\left| 0 \right\rangle _4} + i{\left| 1 \right\rangle _1}{\left| 1 \right\rangle _4})\\
 \otimes ({\left| 0 \right\rangle _2}{\left| 0 \right\rangle _3}{\left| 0 \right\rangle _5}{\left| 0 \right\rangle _6} + i{\left| 1 \right\rangle _2}{\left| 1 \right\rangle _3}{\left| 1 \right\rangle _5}{\left| 1 \right\rangle _6})
\end{array}\)\\
8 & \(\begin{array}{c}
\\
{Y_1}{Y_2}{X_3}{X_4}{X_5}{X_6}{X_7}{X_8} =  - 1\\
{X_1}{Y_2}{Y_3}{X_4}{X_5}{X_6}{X_7}{X_8} =  - 1\\
{X_1}{Y_2}{X_3}{Y_4}{X_5}{X_6}{X_7}{X_8} =  - 1\\
{X_1}{X_2}{X_3}{Y_4}{Y_5}{X_6}{X_7}{X_8} =  - 1\\
{X_1}{X_2}{X_3}{Y_4}{X_5}{Y_6}{X_7}{X_8} =  - 1\\
{X_1}{X_2}{X_3}{X_4}{X_5}{Y_6}{Y_7}{X_8} =  - 1\\
{X_1}{X_2}{X_3}{X_4}{X_5}{Y_6}{X_7}{Y_8} =  - 1\\
{Y_1}{Y_2}{Y_3}{Y_4}{Y_5}{Y_6}{Y_7}{Y_8} =  + 1
\end{array}\) & \(\begin{array}{c}
({\left| 0 \right\rangle _1}{\left| 0 \right\rangle _3}{\left| 0 \right\rangle _4}{\left| 0 \right\rangle _7}{\left| 0 \right\rangle _8} + i{\left| 1 \right\rangle _1}{\left| 1 \right\rangle _3}{\left| 1 \right\rangle _4}{\left| 1 \right\rangle _7}{\left| 1 \right\rangle _8})\\
 \otimes ({\left| 0 \right\rangle _2}{\left| 0 \right\rangle _5}{\left| 0 \right\rangle _6} + i{\left| 1 \right\rangle _2}{\left| 1 \right\rangle _5}{\left| 1 \right\rangle _6})
\end{array}\)\\
\end{tabular}
\end{ruledtabular}
\end{table*}

\providecommand{\noopsort}[1]{}\providecommand{\singleletter}[1]{#1}%

\end{document}